\documentclass{article}
\begin{document}
\def\beq{\begin{equation}}
\def\eeq{\end{equation}}
\newcommand{\bpi}{\mbox{\boldmath $\pi$}}
\def\theequation{\arabic{equation}}
\newcommand{\sgn}{\mathop\mathrm{sgn}}
\renewcommand{\r}{r_0}
\newcommand{\ii}{\mathrm{i}}
\newcommand{\ket}[1]{\left|#1\right\rangle}
\newcommand{\bra}[1]{\left\langle #1\right|}
\begin{center}
{\Large Constant magnetic field and $2d$ non-commutative inverted oscillator}\\
\vspace{0.5 cm}
{\large Stefano Bellucci}\\
\vspace{5mm}
{\it INFN, Laboratori Nazionali di Frascati,
 P.O. Box 13, I-00044 Frascati, Italy}
\end{center}
\begin{center}
{\it e-mail: bellucci@lnf.infn.it}
\end{center}
\begin{abstract}
We consider a two-dimensional non-commutative inverted oscillator in the
presence of a constant magnetic field, coupled to the system in a
``symplectic'' and ``Poisson'' way. We show that it has a discrete
energy spectrum for some value of the magnetic field.
\end{abstract}
\begin{center}
\end{center}
\subsection*{Introduction}
Non-commutative quantum field theories have been
studied intensively during the last several years, owing
to their relationship with M-theory compactifications \cite{cds},
string theory in nontrivial backgrounds \cite{sw} and
quantum Hall effect \cite{hall}
(see e.g. \cite{szabo} for a recent review).
At low energies the one-particle sectors become
relevant, which prompted an interest in the study of
non-commutative quantum mechanics (NCQM) \cite{nair}  - \cite{pAdS}
(for some earlier studies of NCQM see
\cite{Dunne:1990hv} - \cite{Lukierski:1997br}).
Most of the attention was focused on quantum mechanics on
two- and three-dimensional noncommutative spaces. Two-dimensional
NCQM in the presence of a constant
magnetic field was considered on a plane \cite{np,our},
torus \cite{torus}, sphere \cite{np} and pseudosphere
(Lobachewski plane, or $AdS_2$) \cite{iengo, pAdS}.

NCQM on a plane has a critical point,
specified by the vanishing of the dimensionless parameter
  \begin{equation}
 \kappa=1-B\theta\;,
\label{int} \end{equation}
where the system
becomes effectively one-dimensional \cite{np,our}.
Away from the critical point, the rotational properties of the model
become qualitatively dependent on the sign
of $\kappa$: for $\kappa>0$ the system admits an infinite
number of states with a given value of the angular momentum,
while for $\kappa<0$ the number of such states is finite \cite{our}.
From NCQM on a (pseudo)sphere
originate, in some sense, the ``phases'' of planar NCQM \cite{bnsph}:
the ``monopole number'' is defined, in such phases, in a
different way.
In the planar limit the  NCQM on (pseudo)sphere results in the
``non-conventional'', or the so-called ``exotic'' NCQM
\cite{Duval:2000xr}, where the magnetic
field is introduced via ``minimal'', or symplectic coupling.

Notice that the only two-dimensional NCQM with a non-zero potential term that has been solved
explicitly corresponds to a noncommutative circular oscillator in the presence of a magnetic field.
Although this system has been solved for both conventional and exotic coupling of the magnetic field,
its rotational properties have been analyzed only for a conventional coupling of the magnetic field.
The present interest in NCQM was initiated by Chaichian, Sheikh-Jabbari and Tureanu
\cite{Chaichian:2000si}, who calculated the corrections to the hydrogen atom spectrum, which arise
in non-commutative QED. Later on, the three-dimensional noncommutative oscillator with a conventional
coupling of the magnetic field has also been considered and explicitly solved \cite{2dosc}.
Hence, the recent studies of NCQM concentrated mainly on the
systems with either an attractive potential, or without potential.

On the other hand, Ho and Kao \cite{Ho:2001aa}, accurately considering the multi-particle
non-relativistic limit of the noncommutative field theory, found that particles of opposite charges
have opposite non-commutativity.
Hence, in the center-of-mass frame, the ``particle-antiparticle'' system has no noncommutative correction,
in contrast to the system of two identical particles.
Therefore, in the context of a two-particle interpretation, NCQM
with a repulsive potential becomes especially important.\footnote{
In \cite{Ho:2001aa} it was claimed, that ``since proton and electron have opposite charges,
the hydrogen atom has no noncommutative corrections, in spite of the statement
 of \cite{Chaichian:2000si}''. However, the authors of the latter paper argued in their reply \cite{comm}
that this prediction does not work in the case of the hydrogen atom, owing to the composite nature of the proton.}

We recall that, at the present time, the only example of
an exactly solvable
NCQM system with a nonzero potential term is the
oscillator, which adds on to the distinguished role played by the
harmonic oscillator in both theoretical physics and mathematics.\footnote{
A "maximally integrable" isotropic oscillator on the complex
projective space has been recently
introduced in \cite{bnnew}, where the behaviour of the system
in a constant magnetic field
and its supersymmetric generalization on the K\"ahler space
based upon the N=4 mechanics formulated in \cite{bnsup} have been discussed.}
Owing to this reason, in the present work we decided
to examine
the effects of non-commutativity in systems with a repulsive
potential in the presence of a constant magnetic field
(coupled in a conventional and an exotic way), considering the
simple model of the two-dimensional inverted oscillator. For completeness, we shall
present also
the known facts about the non-commutative harmonic oscillator.

\subsection*{Basic properties}

As stated above, we will determine the effects of introducing a
constant magnetic field in the following two cases: the
``conventional'' picture and the ``exotic'' one.
We will also consider, as a by-product, the rotational
properties of the noncommutative oscillator with exotic coupling of the constant magnetic field.
Thus, we consider NCQM with the potential
\begin{equation}
V=\varepsilon\frac{\omega^2 {\bf q}^2}{2}\;,
 \end{equation}
where $\varepsilon=+1$ corresponds to the
harmonic oscillator and $\varepsilon=-1$
to the inverted one.

The two-dimensional non-commutative quantum mechanical
system with an arbitrary central potential in the presence
of a constant magnetic
field $B$ is given by the Hamiltonian \cite{np}
 \begin{equation}
 {{\cal H}}=\frac{{\bf p}^2}{2}+ V({\bf q}^2)\;, \label{h0}
 \end{equation}
with the operators ${\bf p}, {\bf q}$ which obey the
commutation relations
 \begin{equation}\label{xp}
\begin{array}{cc}
 [q_1,q_2]=i\theta\;,\quad [{ q}_\alpha,
{ p}_\beta]=i \delta_{\alpha\beta}\;,\quad
 [p_1,p_2]=iB &{\rm in}\;{\rm the}\;{``conventional''}\;{\rm picture}\\

[q_1,q_2]=\ii{\theta}/{\kappa},
\quad [{ q}_\alpha , { p}_\beta ]=\ii{
\delta_{\alpha\beta}}/{\kappa},
\quad [p_1,p_2]=\ii{B}/{\kappa}&{\rm in}\;{\rm the}\;
{``exotic''}\;{\rm picture},
\end{array}
\end{equation}
where $\alpha,\beta=1,2 $ and the non-commutativity parameter $\theta>0$
has the dimension of ${\rm{\it length}}^2$.

The system can be also represented as follows:
 \begin{equation}
 {\cal H}=\frac{(\bpi+ {\bf q}/\theta )^2}{2} +  V({\bf q}^2)\;,
 \label{8}\end{equation}
where the operators ${\bpi}$ and ${\bf q}$ form the algebra
\beq
[{\pi}_\alpha,{ q}_\beta ] = 0,\;\;
\begin{array}{c}
[\pi_1,\pi_2]=-\ii{\kappa}/\theta, \;\; [q_1,q_2]=i\theta \quad
{\rm ``conventional''}\cr
[\pi_1,\pi_2]=-{\ii}/{\theta},\; \; [q_1,q_2]={\ii}{\theta}/{\kappa}
 \quad {\rm ``exotic''}.
\end{array}\eeq
At the  ``critical point'', i.e. for $\kappa=0$, the system
becomes effectively
one-dimensional \cite{our,Duval:2000xr}
\begin{equation}
[q_1,q_2]=i\theta,\quad {\cal H}^{plane}_0=\left\{
\begin{array}{cc}
{{\bf q}^2}/{2\theta^2}+{ V}({\bf q}^2),&{\rm ``conventional''}\cr
   { V}({\bf q}^2),&{\rm ``exotic''}
\end{array}\right. .
 \label{E0}\end{equation}

Away from the point $\kappa=0$
the  angular momentum of the systems is defined by the operator
\begin{equation}
L=\left\{\begin{array}{cc}
{{\bf q}^2}/{2\theta}-{\theta {\bpi^2}}/{2\kappa}& {\rm ``conventional''}\\
\kappa{{\bf q}^2}/{2\theta}-{\theta {\bpi^2}}/{2}&{\rm ``exotic''}.
\end{array}\right.
\label{ang}\end{equation}
The eigenvalues of the angular momentum operator
are given by the expression
\beq
l=\pm\left( (n_1+1/2)-\sgn{\kappa}\; (n_2+1/2)\right),\quad n_1, n_2=0, 1,...
\label{amp}\eeq
where $(n_1, n_2)$ denote, respectively, the eigenvalues of
the operators
$({\bf q}^2,{\bpi}^2)$ for the ``conventional'' NCQM
and those of the operators $({\bpi}^2,{\bf q}^2)$ for the
``exotic'' one.
In (\ref{amp}) the upper sign corresponds to the ``conventional''
system and the lower sign to the ``exotic'' one.
Hence, the rotational properties of NCQM qualitatively depend
on the sign of $\kappa$.

Let us remind \cite{Duval:2000xr}, that for a non-constant $B$
the Jacobi identities fail in the ``conventional'' model, while
in the ``exotic'' model the Jacobi identities hold for any
$B=A_{[1,2]}$, by definition.
This reflects the different origin of the magnetic field $B$
appearing
in the two models. In the ``conventional'' model, $B$ appears
as the strength of a {\it non-commutative} magnetic field,
 while in the  ``exotic'' model, $B$ appears as
a {\it commutative} magnetic field, obtained by means of
the Seiberg-Witten map from the non-commutative one.

\subsection*{Conventional coupling}

For non-vanishing values of $\kappa$, it is convenient to
introduce the  operators
 \beq
 { a}^{\pm}=\frac{q_1\mp \imath q_2}{\sqrt{2\theta}}\;,
  \quad { b}^{\pm}={\sqrt{\theta}}\frac{\pi_1\mp \imath
 \pi_2}{\sqrt{2|{\kappa}|}}\;,
 \label{nzoperators}
 \eeq
with the following non-zero commutators:
 \beq
 [{ a}^-, { a}^+]=1\;,\quad
 [{ b}^-, { b}^+]= -\sgn \kappa\;.
 \eeq
In terms of the operators in (\ref{nzoperators}),
the angular momentum reads
\beq
 L=({a^+a^- + a^+a^-})/{2}  -
\sgn\kappa \;({b^+b^- + b^+b^- })/{2}\;,
 \label{L}
 \eeq
and the Hamiltonian (\ref{8}) takes the form
\beq
 {\cal H}=\frac{1}{2\theta}\left(|\kappa|(b^+b^-+b^-b^+)-2\imath
 {\sqrt{|\kappa|}}( b^+a^--a^+b^-)+{\cal E}(
a^+a^-+a^-a^+)\right)\;,
\label{H}
\eeq
where we have used the notation
\beq
{\cal E}=1+\varepsilon (\omega\theta )^2\;.
\eeq
Let us introduce  the orthonormal basis in the Hilbert space
consisting of the states
 \beq\label{basis}
 \ket{n_1,n_2}=
\frac{(a^+)^{n_1}(b^{-\sgn\kappa})^{n_2}}{\sqrt{n_1!n_2!}}\ket{0,0}\;,\quad
a^-\ket{0,n_1}=
b^{-\sgn\kappa}\ket{n_2, 0}=0\;,
 \eeq
where $b^{-\sgn\kappa}=b^-$ for
$\kappa>0$, and $b^{-\sgn\kappa}=b^+$ for
$\kappa<0$ .\\

Recall that the eigenvalues of the angular momentum
are given by (\ref{amp}). One can see that the
spectrum essentially depends on the sign
of $\kappa$. Indeed,
the angular momentum $l$ and the occupation number $n_1$
take the values\footnote{
We remind that $n_1$ defines the eigenvalue of the
operator $|{\bf q}|^2/2\theta$ and has the meaning of the
quantized radius of the system $r^2_n=\theta(2n_1+1)$.}

 \beq
  \begin{array}{ccc}
  n_1=0,1,\ldots,&  l=n_1, n_1 +1, \ldots & \mbox{for}\; \kappa<0\;,  \\
 n_1=0,1,\ldots  &  l=-\infty,\ldots,-1,0,\ldots, n_1,
& \mbox{for}\; \kappa>0\;.  \\
 \end{array}
\label{nl} \eeq
At the critical point $\kappa=0$,
the system reduces to the one-dimensional oscillator
with the energy spectrum
\beq
E^{\rm osc}_{(0)n}=
\frac{{\cal E}}{\theta}(n+1/2), \quad n=0,1,2,\ldots
\label{cos}\eeq
Even though we are dealing with an inverted oscillator,
the latter possesses a discrete spectrum,
and gets a ground state when ${\cal E}>0$.

For $\kappa\neq 0$, let us try to
diagonalize the Hamiltonian, by performing an
appropriate (pseudo)unitary transformation
\begin{equation}
 \left(\begin{array}{c}
 {a}\\
 {b}
 \end{array}\right)\to U\cdot
 \left(\begin{array}{c}
 a\\
 b
 \end{array}\right),
\label{t1}\end{equation}
where the matrix $U$ belongs to SU(1,1) for $\kappa>0$, and to
SU(2) for $\kappa<0$,
\begin{equation}\label{u}
  U=\left\{
\begin{array}{c}
  \left(\matrix{
  \cosh \chi e^{\imath \phi}& \sinh\chi e^{\imath \psi}\cr
  \sinh\chi e^{-\imath \psi}&\cosh \chi e^{-\imath \phi}
  }\right),\quad \mbox{for } \kappa>0 \\ {}\\
  \left(\matrix{
  \cos \chi e^{\imath \phi}& \sin\chi e^{\imath \psi}\cr
  -\sin\chi e^{-\imath \psi}&\cos \chi e^{-\imath \phi}
  }\right),\quad \mbox{for } \kappa<0
\end{array}
  \right.
\label{G}\end{equation}
The Hamiltonian becomes diagonal, when $\phi,\psi, \chi$
obey the conditions
\beq
\begin{array}{c}
\cos(\phi+\psi )=0\\
\left\{
  \begin{array}{c}
({\cal E}+\kappa)\sinh 2\chi -2\sqrt{\kappa}\cosh 2\chi \sin(\phi+\psi)=0,
\quad \mbox{for }
  \kappa>0\\
 ({\cal E}+\kappa)\sin 2\chi +2\sqrt{-\kappa}\cos 2\chi \sin(\phi+\psi)=0,
\quad \mbox{for }
  \kappa<0\\
  \end{array}\right. .
\end{array}
\label{t2}\eeq

The diagonalized Hamiltonian should take the form
\begin{equation}\label{osc}
 {\cal H}_{\rm osc}=
 \frac{1}{2}\omega_-(b^+b^- +b^-b^+)+
\frac{1}{2}\omega_+(a^+a^-+
 a^-a^+)\;,
\label{t3}\end{equation}
where
\begin{equation}
 {2\theta\omega_\pm}=\left\{
 \begin{array}{c}
  \pm(\kappa-{\cal E})+({\cal E}+
\kappa)\cosh 2\chi-2\sqrt{\kappa}\sinh 2\chi
\sin(\phi+\psi) \; ,\quad \mbox{for }
  \kappa>0\\
  ({\cal E}-\kappa)\pm\left[
  ({\cal E}+\kappa)\cos 2\chi
 -2\sqrt{-\kappa}\sin 2\chi \sin(\phi+\psi)\right]\; ,
\quad \mbox{for }  \kappa<0 \; .
 \end{array}\right.
\label{250}\end{equation}

{\bf \large ``Conventional'' harmonic oscillator.}
In the case of the
harmonic oscillator the above equations have a solution for any
$\kappa$. After some work one gets
\begin{equation}
 {2\theta\omega_\pm}=\left\{
 \begin{array}{c}
  \pm({\cal E}-\kappa)+\sqrt{({\cal E}+
\kappa)^2-4\kappa},\quad \mbox{for }
  \kappa>0\; ,\\
  ({\cal E}-\kappa)\pm
  \sqrt{({\cal E}+\kappa)^2-4\kappa},\quad \mbox{for }
  \kappa<0  \; .
 \end{array}\right.
\end{equation}
Hence, the energy spectrum of the ``conventional'' oscillator
  takes the form
\begin{equation}\label{oscspectrum}
\begin{array}{cc}
 E^{\rm osc}_{n_a,n_b}=&
 \omega_+(n_a+{1}/{2})
 +\omega_- (n_b+{1}/{2})=\cr
\; &=\left[
 (\sqrt{({\cal E}+\kappa)^2-4\kappa}(n_1+1/2)-
(\sqrt{({\cal E}+\kappa)^2-
4\kappa} +\kappa-{\cal E})l \right]/\theta\; .
\end{array}
\end{equation}
Since the transformation (\ref{u}) belongs to the
symmetry group of the rotational momentum $L$, the magnetic number
is given by the same equation as above, i.e. (\ref{amp}).
It can be seen that the expressions
(\ref{nl}) arise, in the case of the harmonic oscillator,
from the requirement of the
positivity of the energy spectrum.

There exists an ``isotropic point",
 ${\cal E}=\kappa \;>1$,
where the frequencies become equal to each other
$$\omega^{\rm{isotr}}_\pm=
\omega\sqrt{1+(\omega\theta )^2}\;,$$
and the system has the symmetry of the ordinary circular oscillator.

In the commutative limit, i.e. for $\theta\to 0$, the
effective frequencies read
\begin{equation}
\omega^0_\pm= \pm{B}/{2}+\sqrt{\omega^2+
{B}^2/{4} }\;.
\end{equation}
In the case
of the Landau problem, ${\cal E}=1$ (or, equivalently, $\omega=0$),
one of the frequencies vanishes and the spectrum reads
$$
E_n=|B|(n+\frac 12),\quad
l=n_1-\sgn\kappa n_2,\quad n=\left\{
\begin{array}{ccc}
n_2=0,1,\ldots &{\rm for}&\kappa>0\\
n_1=0,1,\ldots &{\rm for }&\kappa<0\;.
\end{array}\right.
$$
Hence, although the energy spectrum of the
Landau problem is independent from the non-commutativity parameter,
its expression in terms of the angular momentum essentially depends
on $\sgn\kappa$.\\

{\bf \large ``Conventional'' inverted oscillator}
When we deal with the ``conventional'' inverted oscillator,
 $\varepsilon=-1$,  the equations (\ref{250})
have a solution for any $\kappa<0$, as well as for the values
$\kappa>0$ which satisfy the condition
\begin{equation}
({\cal E}+
\kappa)^2-4\kappa \geq 0\quad .
\end{equation}
Hence, the non-commutativity of the plane,
together with the presence of a magnetic field, yield different
``regimes'' in the ``conventional'' inverted oscillator:
\begin{itemize}
\item $\kappa=0$:
the spectrum of the inverted oscillator is given by the expression
(\ref{cos}),
where ${\cal E} < 1$ .
Hence, when ${\cal E}>0$, the inverted oscillator transmutes into
a one-dimensional harmonic one, and the system possesses a ground state.
For ${\cal E}=0$ the energy of the system vanishes.

\item  $\kappa<0 $: the Hamiltonian
of the inverted oscillator can be diagonalized by
a $SU(2)$
transformation.
The resulting system is again given by (\ref{t3}), where
$\omega_+ >0$, $\omega_-<0$.
\item $\kappa>0$, $(\kappa +{\cal E})^2 > 4\kappa $ (or, equivalently,
$\kappa \in [\; 0,\;
 \left(1-{\omega\theta}\right)^2 ]\;\cup\; [
\left(1+{\omega\theta}\right)^2,\infty ]$ .\\
In this case, we can diagonalize  the Hamiltonian
by an appropriate $SU(1.1)$ transformation.
The energy spectrum is defined by  by (\ref{t3}), where
$$
\begin{array}{cc}
\omega_+>0,\quad \omega_-<0 &{\rm for}\; {\cal E}>\kappa \\
\omega_+<0,\quad \omega_->0 &{\rm for}\; {\cal E}<\kappa\;.
\end{array}
$$
Hence, although the spectrum is discrete, the system has no ground state.
This regime also appears in the absence of a magnetic field, i.e. for $B=0$,
when ${\cal E}<-3$.
In this case the ``frequencies'' are
$\omega_\pm=0,\;\; 2(\omega\theta)^2$.

\item  $\kappa>0$, $(\kappa +{\cal E})^2 \leq 4\kappa $ (or, equivalently,
$\kappa \in [ \left(1-{\omega\theta}\right)^2\;,
\left(1+{\omega\theta}\right)^2\; ]$).\\
In this case we cannot diagonalize the Hamiltonian,
neither by $SU(1.1)$, nor by $SU(2)$ transformations.
This indicates that in the given regime
the system possesses a smooth energy spectrum.
Notice that this is the only regime which has commutative limit.
 \end{itemize}
So, considering the simplest system with a
repulsive potential, we find
that the non-commutativity of the coordinates, together with the
presence of a non-vanishing magnetic field,
essentially change its initial properties.\\

\subsection*{ ``Exotic'' coupling}

Here we must distinguish, once more, between two different cases.\\

{\bf\large ``Exotic'' harmonic  oscillator.}
The spectrum of the ``exotic'' oscillator, away from the point $\kappa=0$,
can be obtained in a way similar to that described above.
For this purpose we should make, in the above derivation, the following
replacements:
\beq
\bpi\to - {\bf q}/\theta,\quad {\bf q}/\theta\to -\bpi,\quad
 \kappa\to 1/\kappa\;.
\eeq
Thus, away from the point $\kappa=0$, the spectrum of the ``exotic''
oscillator reads
\begin{equation}\label{oscspectru }
 E^{\rm osc}_{n_1,n_2}=
 \omega_-(n_1+{1}/{2})
 +\omega_+ (n_2+{1}/{2})\;,
\eeq
where
\begin{equation}
 {2\theta\omega_\pm}=\left\{
 \begin{array}{c}
  \pm({\cal E}-1/\kappa)+\sqrt{({\cal E}+
1/\kappa)^2-4/\kappa},\quad \mbox{for }
  \kappa>0\\
  ({\cal E}-1/\kappa)\pm
  \sqrt{({\cal E}+1/\kappa)^2-4/\kappa},\quad \mbox{for }
  \kappa<0 \; .
 \end{array}\right.
\end{equation}
Since the transformation (\ref{u}) belongs to the
symmetry group of the angular momentum $L$, the eigenvalues of
the latter operator are given by (\ref{amp}).
As in the case of the ``conventional'' oscillator, the expressions
(\ref{nl}) arise, also in the case of the ``exotic'' oscillator,
from the requirement of the
positivity of the energy spectrum. The ``isotropy point" of
the ``exotic''  oscillator is defined by the expression
\beq
 {\cal E}=1/\kappa,\quad \kappa>0\; .
 \eeq

{\bf \large ``Exotic'' inverted oscillator.}
The  regimes of the ``exotic''
inverted oscillator can be found
in the same way, as the regimes of the
inverted ``conventional'' oscillator.
One has
\begin{itemize}
\item  $\kappa<0 $: the Hamiltonian
of the inverted oscillator can be diagonalized by
a $SU(2)$
transformation.
\item $\kappa>0$, $(1/\kappa +{\cal E})^2 > 4/\kappa $ (or, equivalently,
$\kappa \in [\; 0,\;
 \left(1+{\omega\theta}\right)^{-2} ]\;\cup\; [
\left(1-{\omega\theta}\right)^{-2},\infty ]$ .\\
In this case we can diagonalize  the Hamiltonian
by an appropriate $SU(1.1)$ transformation.
The energy spectrum is discrete, however the system has no ground state.
\item  $\kappa>0$, $(1/\kappa +{\cal E})^2 \leq 4/\kappa $ (or, equivalently,
$\kappa \in [ \left(1+{\omega\theta}\right)^{-2}\;,
\left(1-{\omega\theta}\right)^{-2}\; ]$).\\
In this  case we cannot diagonalize the Hamiltonian,
neither by $SU(1.1)$, nor by $SU(2)$ transformations.
The system has a smooth energy spectrum.
\end{itemize}

\subsection*{Discussion}

We considered quantum mechanics with a repulsive oscillator potential
on a non-commutative plane, interacting with a constant magnetic
field. This system could be interpreted as a two-particle system in a
center-of-mass frame. We found that non-commutativity, combined with the presence
of a constant magnetic field, generates a special regime, where the system gets
a discrete spectrum. The qualitative behaviour of the system is the same for
both types of coupling of the magnetic field, i.e. symplectic and Poisson ones
(which we called, respectively, ``conventional and ``exotic'').

It should be quite interesting to take into consideration similar non-commutative
systems on two- and higher-dimensional spheres, as well as to study
non-commutativity effects on supersymmetric quantum mechanics.
The appropriate {\it commutative} candidate, namely a (super)oscillator, in
a constant magnetic field, on complex projective spaces,
is constructed in \cite{bnnew}.\\

It can be seen from eq. (\ref{E0}) that, at the
critical point, the ``conventional'' NCQM gets a ground state
not only for attractive potentials, but for repulsive potentials
as well, when the latter are long-range distance ones, namely
\beq
{2\theta^2}\frac{dV(q^2)}{dq^2}>-1\;.
\eeq
This class includes the physically important
cases of the logarithmic potential
$
V       =-{\gamma^2}\log{{\bf q}^{2}}/{2\theta}\;
$
and the Coulomb $V= {\gamma^2}/{|{\bf q}|}$ one,
as well as a simplest repulsive potential, i.e. that describing
an inverted oscillator.\\

\subsection*{Acknowledgments}

I would like to thank Armen Nersessian for useful discussions.
This work was supported in part
by the European Community's Human Potential
Programme under the contract HPRN-CT-2000-00131 Quantum Spacetime,
the INTAS-00-0254 grant, the NATO Collaborative Linkage Grant PST.CLG.979389 and the Iniziativa
Specifica MI12 of the INFN Commissione Nazionale IV.

\end{document}